\documentclass[twocolumn,aps,pre,superscriptaddress,floats,floatfix]{revtex4-1}
\usepackage{amsmath}
\usepackage{amssymb}
\usepackage{epsfig}
\usepackage{subfigure}
\usepackage{graphicx}
\usepackage{bm}
\usepackage{xcolor}
\usepackage{float}
\usepackage{sidecap}
\usepackage{wrapfig}
\def\la{\left\langle}
\def\ra{\right\rangle}

\begin{document}
\title {Revisiting the SABRA Model: Statics and Dynamics}
\author{Rithwik Tom}
\email{rithwikiisc@gmail.com}
\affiliation{Department of Physics, Indian Institute of Science, Bangalore 560012, India}
\altaffiliation{Presently: Department of Physics, Carnegie Mellon University, 5000 Forbes Avenue, Pittsburgh PA 15213-3890, USA}
\author{Samriddhi Sankar Ray}
\email{samriddhisankarray@gmail.com}
\affiliation{International Centre for Theoretical Sciences, Tata Institute of Fundamental Research, Bangalore 560089, India}
\begin{abstract}
We revisit the two-dimensional SABRA model, in the light of recent results 
of Frisch {\it et al.} [Phys. Rev. Lett. {\bf 108}, 074501 (2012)] and examine, systematically,
the interplay between equilibrium states and cascade (turbulent) solutions, characterised 
by a single parameter $b$, via 
equal-time and time-dependent structure functions. We calculate the static and 
dynamic exponents across the equipartition as well as turbulent regimes which 
are consistent with earlier studies. Our results indicate the absence of a sharp 
transition from equipartition to turbulent states. Indeed, we find that the SABRA model 
mimics true two-dimensional turbulence only asymptotically as $b\to-2$.
\end{abstract}
\maketitle

The search for a single, unifying framework to describe turbulent flows remains 
one of the most important challenges in classical physics. Given the essential nature of turbulent flows and the solutions of the
Navier-Stokes equation, significant progress has been made in our understanding
through methods of theoretical physics and, in particular, statistical physics.
Since the pioneering work of Kolmogorov~\cite{k41}, and typically working
within the homogeneous and isotropic idealisation, significant advances have
been made in understanding, for example, the nature of correlations in such
flows~\cite{frischbook,falcormp,pandit-review}. Indeed, the universality of power-laws
in such correlation functions, reminiscent of critical
phenomena, raises the possibility of using the language of
statistical mechanics and renormalization group to gain a microscopic
understanding of turbulence and the underlying Navier-Stokes equation.

Despite this, it has not always been possible to adapt tools of statistical
mechanics with the same degree of success.  An inherent contradiction between the use of
standards statistical physics approaches and real, turbulent flows is best
captured in the work of Hopf~\cite{hopf} and Lee~\cite{lee}. By using the
Hamiltonian system of the incompressible Euler equation (with viscosity $\nu =
0$) by projecting, via a Galerkin projection, to a finite-dimensional system,
it was shown~\cite{hopf,lee} that the solution thermalises leading to an energy
spectrum $E(k) \sim k^2$ in stark contrast to the famous Kolmogorov scaling
$E(k) \sim k^{-5/3}$.  Indeed, since the remarkable discovery of Cichowlas {\it et
al.}~\cite{brachet05}, through state-of-the-art direct numerical simulations (DNSs)
that although the truncated Euler equation will eventually lead to
equipartition, there are long-lived transient states which are partially
thermalised admitting both Navier-Stokes-like and equilibrium solutions.  We
now have a fairly good understanding of the origins of thermalised
states~\cite{ray11} (see also~\cite{bannerjee14,ray-review,divya17}), mediated by structures
called {\it tygers} in Ref.~\cite{ray11}, it is often the partially thermalised
regime which has proved important in
understanding~\cite{bannerjee14,frisch08,frisch13} more practical aspects of
turbulence, such as the ubiquitous bottleneck effect~\cite{bottleneck}

A major breakthrough in understanding the relationship between
equilibrium statistical mechanics and turbulence was made by L'vov {\it al.}~\cite{lvov02} who 
showed the existence of flux-less,
equilibrium solutions, which coincide with the Kolmogorov scaling at a critical
dimension. More recently this was checked numerically by Frisch {\it et
al.}~\cite{frisch12} via the method of Fourier decimation introduced in the same
paper (see, also Refs.~\cite{ray-review,lvov02}). Subsequently
the issue of possible equilibria solutions and its connection to intermittency
has been studied extensively in the last three years in a series of papers both
for the Navier-Stokes equation~\cite{luca15,luca16,michele17} as well as the
analytically more tractable Burgers equation~\cite{michele16}.

This large body of work in the last couple of years have renewed interest in
the problem of equilibrium solutions in equations of hydrodynamics and their
possible implications for turbulence. However so far the investigations have
been confined to two-point, equal-time correlation functions and the issue of
dynamics of such systems has been left largely unexplored. This is because
examining time-dependent correlation functions, either analytically or through
direct numerical simulations of the (decimated) Navier-Stokes equation, is
still a major challenge. Indeed, results for time-dependent structure functions
and the variety of time-scales (leading to multiscaling) associated with them
in fully-developed turbulence have been obtained numerically mainly in reduced
models~\cite{mitra04,mitra05,raynjp,rayepjb}, such as the GOY shell model, with
far fewer results from DNSs of the Navier-Stokes equation~\cite{luca11,rayprl11}
with theoretical underpinnings in the Parisi-Frisch multifractal
formalism~\cite{parisifrisch}. 
 
We adopt the spirit of previous studies on dynamic scaling in turbulence 
by resorting to numerically tractable shell models and report the first results 
on the dynamics of a turbulent systems near {\it equilibrium}. We thus 
revisit the {\it two-dimensional} SABRA model and examine, systematically,
the interplay between equilibrium states and cascade (turbulent) solutions as a
function of a single parameter $b$. This work builds on previous studies by
Ditlevsen and Mogensen~\cite{mogensen} and Gilbert, {\it et
al.}~\cite{gilbert2002}. In particular, their studies had shown a rich phase
diagram in the solution to the two-dimensional SABRA model with cross-overs to
turbulent and equilibrium solutions. We now closely examine this cross-over, 
in the light of what we have learnt from decimated systems, for the equal-time 
structure functions before understanding the dynamics and time-scales associated 
with such equilibrium states. The use of shell models to study dynamics have, apart from being a surrogate 
because of the problems of DNSs, several advantages. Chief amongst these are the 
fact that it allows us to resolve a huge range of scales, inaccessible to modern 
day computers for DNSs, and also, naturally (as we explain below) eliminate sweeping 
which can lead to trivial scaling~\cite{luca-review,pandit-review}.

In this paper we work with the SABRA shell model~\cite{sabra}:
\begin{eqnarray}
\left[\frac{d}{dt} + \nu k_n^2 + \mu k_n^{-2}\right]u_n &=& \imath\bigg{[}a k_{n+1}u_{n+1}^{\ast}u_{n+2} \nonumber \\ 
&+&bk_nu_{n-1}^{\ast}u_{n+1} \nonumber \\ &-&ck_nu_{n-1}u_{n-2} \bigg{]} + f_n.
\label{sabra}
\end{eqnarray}

This set of coupled ordinary differential equations are augmented by the boundary conditions are 
$u_{-1} = u_0 = 0; u_{N+1} = u_{N+2} = 0$, where $N$ is the maximum number of shells used. 
The scalar wave vectors are conventionally written in the form $k_n = k_0 \lambda^n $; we use typical values 
$\lambda = 2$ and $k_0 = 1/16$. Since we are studying the two-dimensional model, the coefficients $a = 1$, $b$, and $c$, 
satisfying the constraint $a + b + c = 0$, are chosen to conserves the shell-model analogues of energy
\begin{equation}
E = \frac{1}{2}\sum_{n=1}^{N} |u_n|^2
\end{equation}
and a generalised enstrophy
\begin{equation}
\Omega = \frac{1}{2}\sum_{n=1}^{N} k_n^{\alpha}|u_n|^2
\end{equation}
in the inviscid, unforced limit for $-2 < b < -1$, with $\alpha =  -\log_{\lambda}|-b-1|$~\cite{goy,ditlevsenbook,bohrbook}. 
We use an external force $f_n$ to drive the
system to a (non-equilibrium) steady state. All measurements of equal-time and
time-dependent structure functions are made in this steady state. 
The SABRA model can thus be studied by varying a single free parameter, in this case, $b$ continuously 
since the $a = 1$ and $c = -1 - b$. .

\begin{figure}
\begin{center}
\includegraphics[scale=0.16]{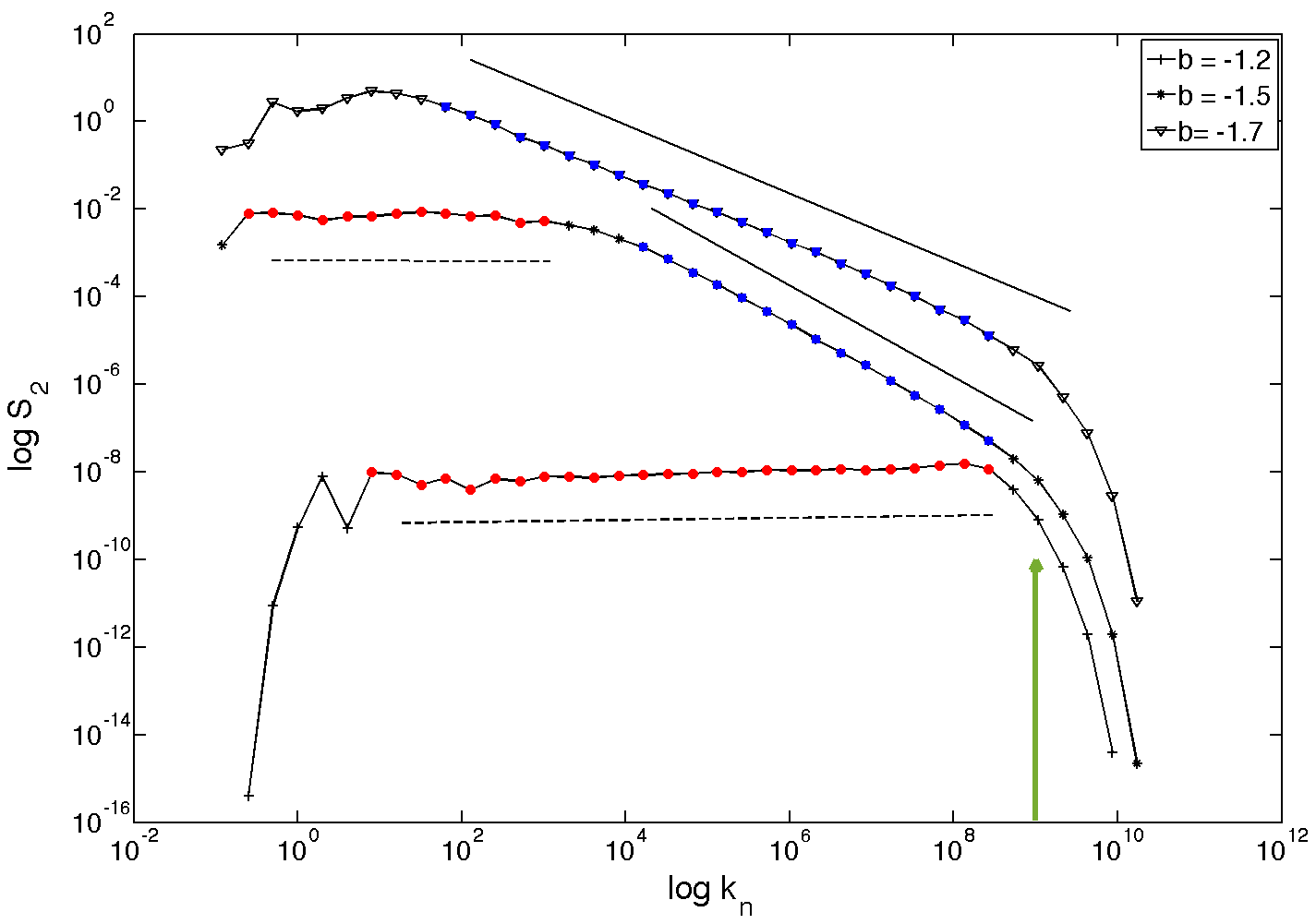}
\end{center}
\caption{(color online) Representative log-log plots of the second-order equal-time structure functions $S_2$ versus 
$k_n$ from run {\bf R1} ($N = 43$) for $b = -1.2$ (lowest curve), $b = -1.5$ (middle curve), and $b = -1.7$ (top curve). 
The forcing scale is indicated by the green vertical arrow. For the largest value of $b$ we see that the scaling regime is a 
plateau (energy equipartition), 
shown by the red circles and indicated by the horizontal, dashed line; for the smallest value of $b$, we see a clear power-law 
scaling, with a non-zero exponent (enstrophy equipartition), as shown by the blue triangles and the thick black line as a guide to the eye. For an 
intermediate value of $b = -1.5$, there are clearly two scaling regimes, namely the flat plateau (shown by the red circles and 
the dashed line) and a second power-law with a non-zero scaling exponent (shown by the blue triangles and the thick black line). 
The curves have been shifted arbitrarily for clarity of presentation. A summary of the various exponents and their dependence on 
$b$ is shown in Fig.~\ref{zeta}.} 
\label{Sp2}
\end{figure}

\begin{figure*}
\begin{center}
\includegraphics[scale=0.16]{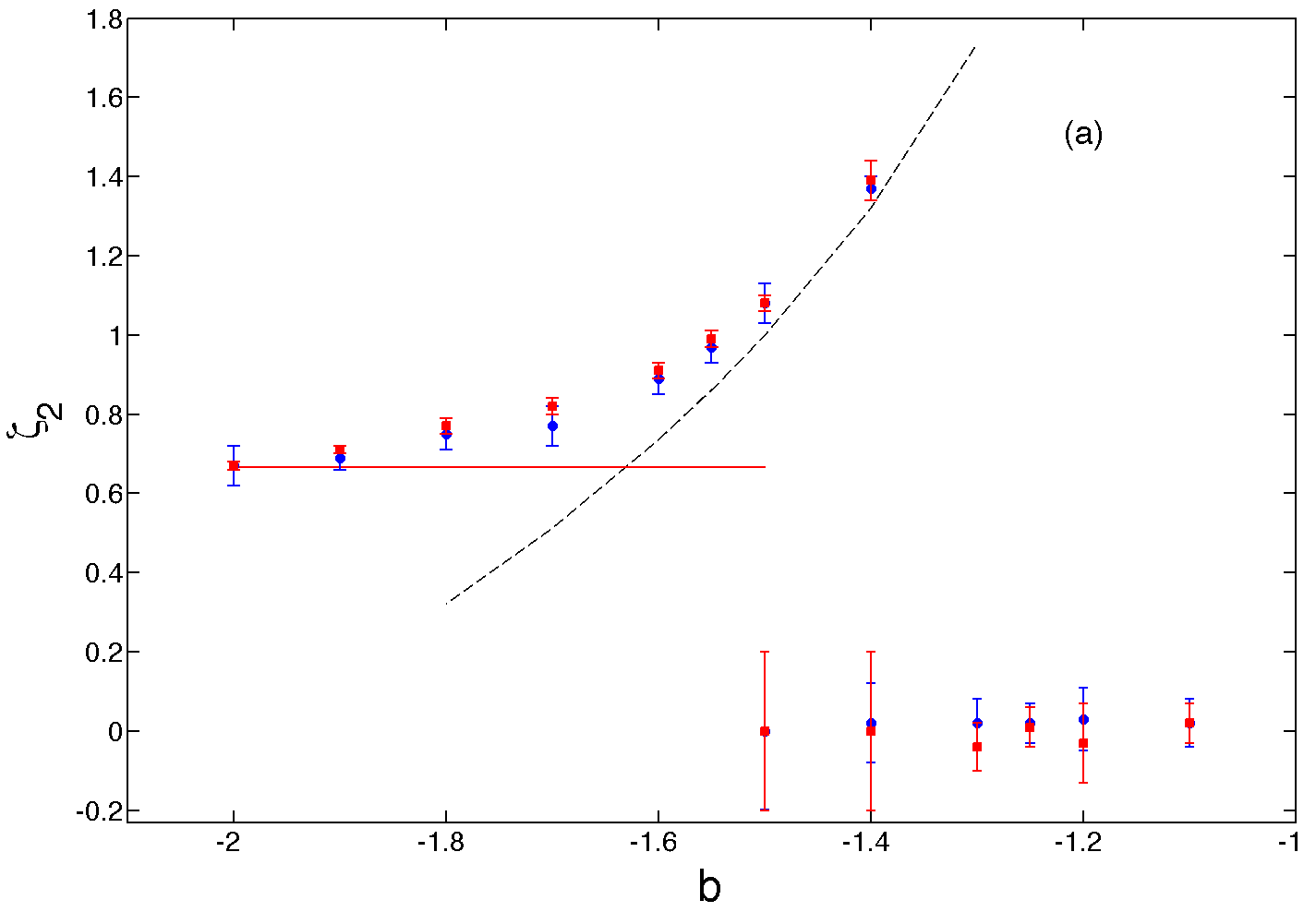}
\includegraphics[scale=0.16]{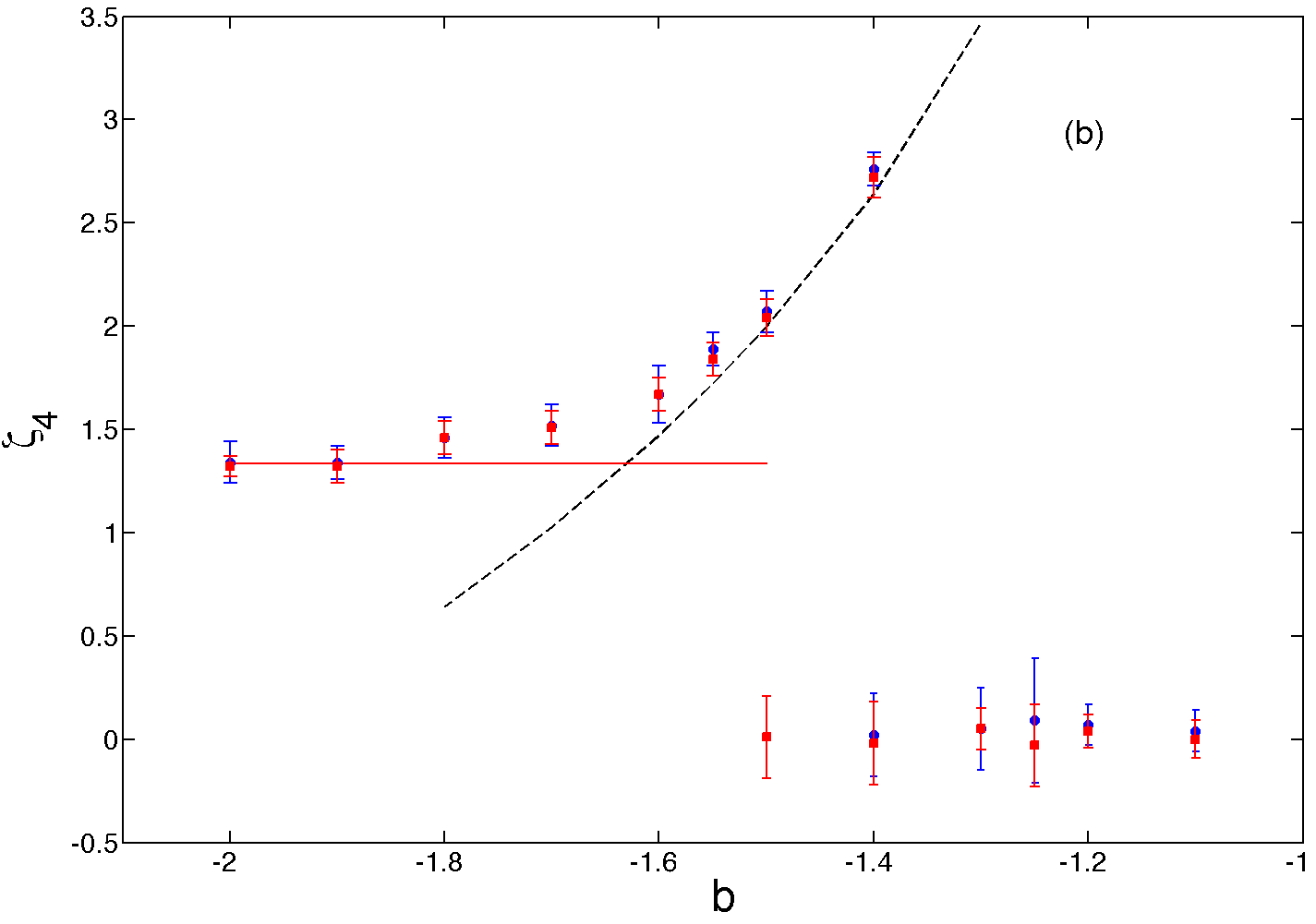}
\end{center}
\caption{(color online) Equal-time exponents (a) $\zeta_2$ and (b) $\zeta_4$ as a function of the parameter $b$ from our numerical 
simulations {\bf R1} and {\bf R2} of the SABRA model. The various exponents are extracted 
from log-log plots of the equal-time structure functions as illustrated in Fig.~\ref{Sp2}. 
The lower branch correspond to an energy equipartition regime, yielding 
an exponent $\zeta_2 = \zeta_4 = 0$ (within error-bars) disappear close to the conjectured critical value $b_c$.  
The upper branch correspond to an enstrophy equipartition regime before converging to the turbulent, Kolmogorov-like 
solution $\zeta_2 = 2/3$  and $\zeta_4 = 4/3$ (indicated by the horizontal red dot-dashed line in panel). The black, dashed line in 
panel denotes the exponent predicted from the enstrophy partition regime, namely $\zeta_2 = \alpha = -\log_{\lambda}|-b-1|$ and 
$\zeta_4 = 2\zeta_2$ for $b>b_c$ and 
for shells larger than $n_c$ but smaller than $n_f$. We note that, within 
error-bars, the measured exponents are slightly higher than those obtained theoretically and two-dimensional, turbulent behaviour is only approached 
asymptotically as $b \to -2$. The exponents obtained from the two sets of simulations, shown as blue circles ({\bf R1}) and red squares 
({\bf R2}) are in agreement with each other within error-bars.}
\label{zeta}
\end{figure*}

In Ref.~\cite{gilbert2002}, the authors showed through a mixture of numerics
and theory that the as a function of $b$ the solution of the SABRA model shows
a phase diagram with not only the inverse and direct cascade regimes, on either
side of the forcing shell $n_f$, but also energy and enstrophy equipartition
phases on either side of a cross-over shell number (which itself is a function
$b$) $n_c < n_f$ with a phase boundary separating which disappears at $b = b_c
= -a(1+\lambda^{-2/3}) \approx -1.63$.  For values $-2 < b < b_c$, such
equilibrium solutions disappear leaving only the dual cascade picture typical
of two-dimensional turbulence.  It is this rich phase diagram which makes the
two-dimensional SABRA model an ideal candidate to study the dynamics in the
interplay between turbulence (cascade) and fluxless solutions.  In this paper,
we revisit the two-dimensional SABRA model and systematically study its
statics and dynamics for the full range of $-2 < b < -1$.

We perform two different sets of simulations (runs {\bf R1} and {\bf R2}) 
of Eq.~(\ref{sabra}); in each case we use 11 different values of $-2 < b < -1$ with the
total number of shells $N = 43$ ({\bf R1}) and $N = 28$ ({\bf R2}). We use, 
in addition to the normal viscosity, a hypo-viscous term $\mu k_n^{-2}$ to drain the energy at lower
shell numbers which accumulate because of inverse cascade ~\cite{foot1}.  We
use initial conditions $u_n^0 = k_n^{1/2} e^{\imath \vartheta_n},$ for $n = 1,
2$, and $u_n^0 = k_n^{1/2}e^{{-k_n}^2} e^{\imath \vartheta_n},$ for $3 \leq n
\leq N$; $\vartheta_n$ is a random angle distributed uniformly between $0$ and
$2\pi$.  We drive the system to steady state by using a deterministic forcing
(a) $f = (1 + \imath)\times 5\times 10^{-3}$ on shell $n_f = 22$ ({\bf R2}) and (b) 
$f= (1 + \imath)\times 8\times 10^{-3}$ on shell $n_f = 34$
({\bf R1}).  We use a slaved,
Adams-Bashforth scheme~\cite{pisarenko,dhar} to integrate Eq.~(\ref{sabra}) by
using a time step (a) $\delta t = 10^{-4}$, $\nu = 10^{-7}$, and $\mu = 10^{-2}$ ({\bf R2}) 
and (b) $\delta t = 10^{-7}$, $\nu = 10^{-14}$, and $\mu = 3\times10^{-4}$  ({\bf R1}).
Given the delicate measurements, it is important to rule out any finite-size (in the
scaling range) effects; hence we use these two different-sized simulations and 
find our results from {\bf R1} and {\bf R2}, which we report below, to agree with each 
other within error-bars.

We begin with the $p$-th order, equal-time structure function and the associated 
equal-time exponent $\zeta_p$. For the shell model this is defined via
\begin{equation}
S_p(k_n) \equiv \la [u_n(t)u^{\ast}_n(t)]^{p/2} \ra \sim k_n^{-\zeta_p},
\label{eqstfun}
\end{equation}
where $\la \cdot \ra$ denotes an average over time in the steady state. 

We compute the $p$-th order structure function \eqref{eqstfun} by averaging, in
the steady state, over a time window $\Delta t \gg 1$. We choose 50 such
statistically independent time windows and thence obtain 50 values of the
equal-time scaling exponents. We quote the mean of these exponents as $\zeta_p$
and their standard deviation is a measure of the error-bar on them. 

For the two-dimensional SABRA model, for $b > b_c$ and $n < n_f$, Ditlevsen and
Mogensen ~\cite{mogensen} conjectured that the second-order structure function should show 
two scaling regimes with different scaling exponents. Thus for lower wavenumber $\zeta_2 =
0$ for $n < n_c$ due to energy equipartition and at higher wavenumbers $\zeta_2 = \alpha$ for $n_c <
n < n_f$ because of enstrophy equipartition. On the other hand, for $b < b_c$,
similar arguments ~\cite{gilbert2002} leads one to the conclusion that  a single scaling 
range emerges with $\zeta_2 = 2/3$ for $n < n_f$ because of the inverse cascade of two-dimensional turbulence.

\begin{figure*}
\begin{center}
\hspace*{-1.0cm}
\includegraphics[scale=0.16]{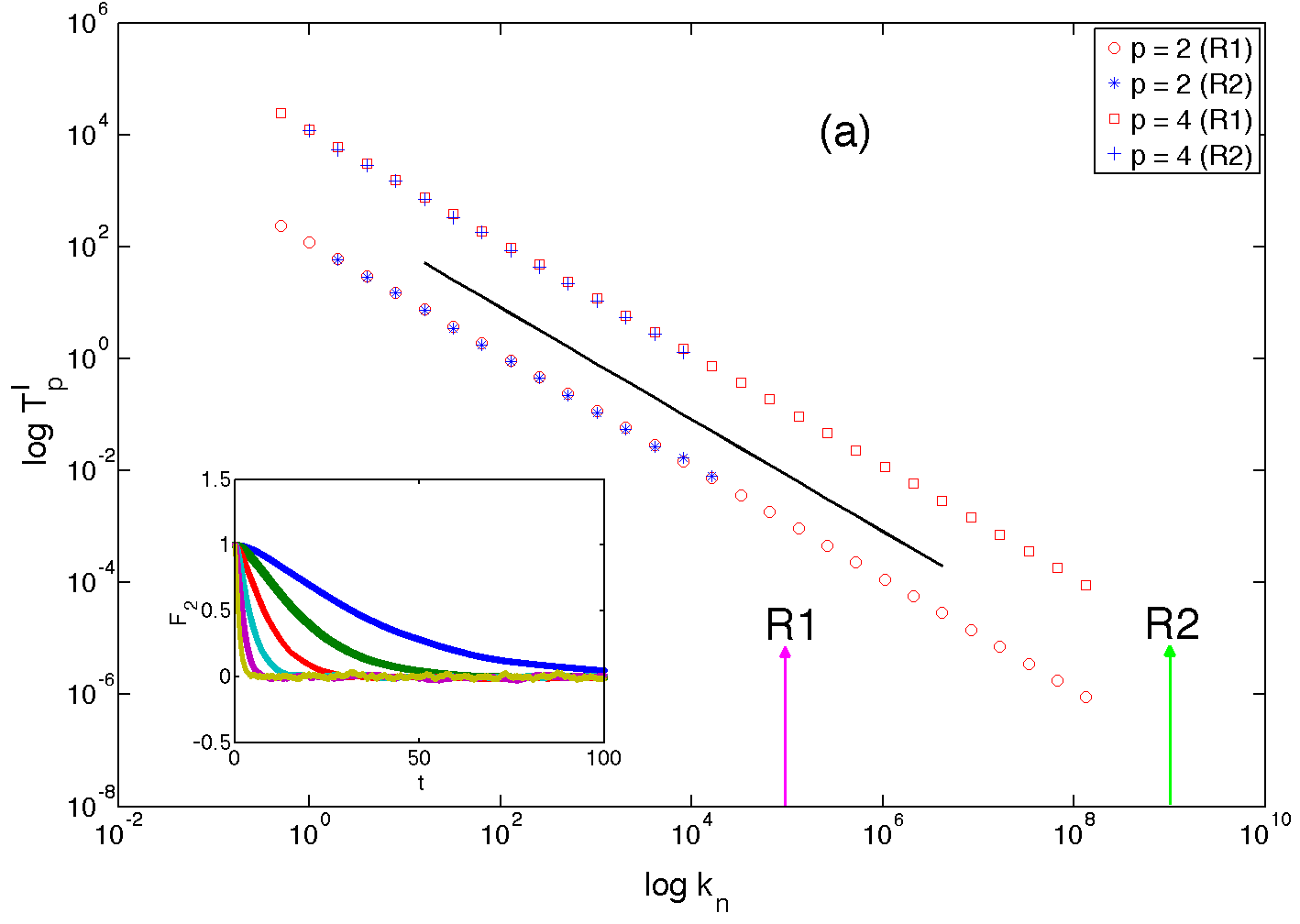}
\includegraphics[scale=0.155]{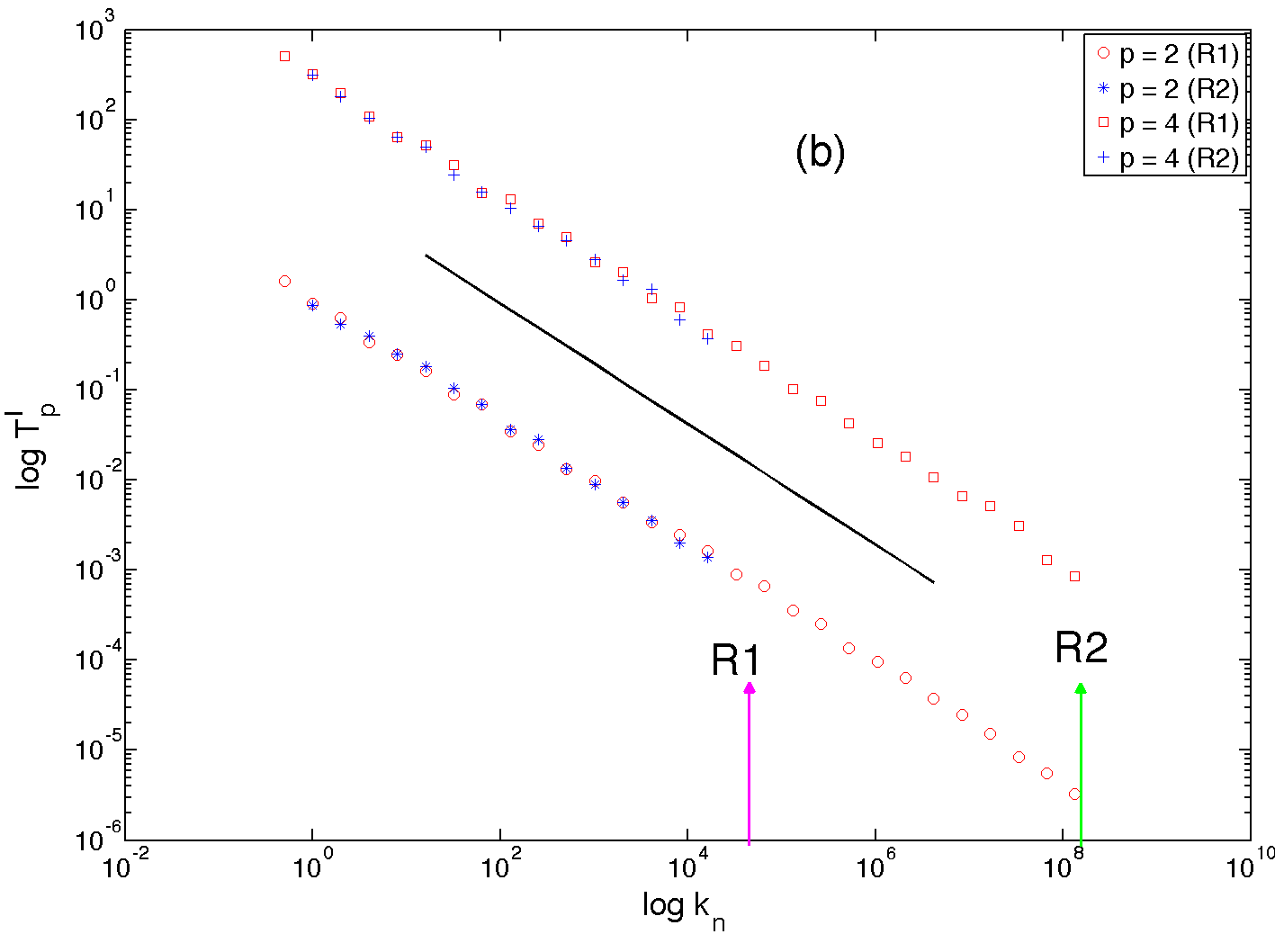}
\end{center}
\caption{(color online) Representative log-log plots, from runs {\bf R1} (red $\circ$ and $\square$) and {\bf R2} (blue * and +), 
of the second and fourth-order integral time scales versus $k_n$, for extremal values of $b$, namely (a) $b = -1.2$ and (b) $b = -1.9$.
The black, dashed line corresponds to our theoretical prediction in the (a) energy equipartition ($k_n^{-1}$) and 
(b) turbulent, Kolmogorov-like ($k_n^{-2/3}$) regimes. The upper set of curves in both panels are for $T^I_4$ and the lower set 
of curves for $T^I_2$. The forcing scales are indicated by the vertical pink and green arrows as labelled in the figure. In the inset of 
panel (a), we show a representative plot of the time-dependent, second-order structure  function $F_2$, from where the time-scales 
are extracted, for several different wavenumbers 
(inner curves for the larger shell numbers and outer ones for the smaller) as an inset in panel (a). The transition from the equipartition regimes to the 
turbulent ones is illustrated in Fig.~\ref{T2}.}
\label{T2-asym}
\end{figure*}

In Fig.~\ref{Sp2} we show representative plots of the second-order, equal-time structure function 
for different values of $b$. For values of $b > b_c$, we find a single plateau ($\zeta_2 = 0$) shown by 
red circles consistent with earlier predictions. For $b < b_c$, a different scaling range, shown with blue 
triangles, with a power-law exponent $\zeta_2 \neq 0$. For values of $b$ close to and around the 
critical $b_c$, we find the existence of both these scaling ranges as is clearly seen in Fig.~\ref{Sp2}.

It is also useful to keep in mind, that a similar conclusion 
can be drawn by measuring the fluxes as was shown in Ref.~\cite{gilbert2002}. 

In Fig.~\ref{zeta} we show plots of the exponents (a) $\zeta_2$ and (b)
$\zeta_4$ for 12 different values of $b$ from runs {\bf R1} (red squares) and {\bf R2} (blue circles). 
For large values of $-1.3 \lesssim b
\lesssim -1.0$, the scaling range exhibits only the energy equipartition range
($n_c \approx n_f$) yielding a single exponent, within error-bars, $\zeta_2 =
0.0$. For lower values $b_c \lesssim b \lesssim -1.4$, the second-order
structure functions show dual scaling (for $n < n_f$): $\zeta_2 = 0$ for
$n<n_c$ and $\zeta_2 = \alpha$ for $n_f < n < n_c$. In  Fig.~\ref{zeta}(a) we see
the two branches in the values of $\zeta_2$ in the range $b_c \lesssim b
\lesssim -1.4$. The lower branch $\zeta_2 = 0$ is consistent with the energy
equipartition prediction. The black, dashed line corresponds to enstrophy
equipartition $\alpha$: However, we find that the measured exponent $\zeta_2$ in
the second scaling range although close to the enstrophy equipartition
prediction is nevertheless marginally larger than $\alpha$. For values of $b
\gtrapprox b_c$, the scaling of the energy equipartition disappears leaving
only the second scaling regime $\zeta_2 \approx \alpha$. We finally note that 
this single branch of $\zeta_2$ asymptotes to the value 2/3 (denoted by the red, horizontal 
dot-dashed line) as $b \to -2$. In Fig.~\ref{zeta}(b) we show an analogous plot for the 
fourth-order exponent $\zeta_4$ which shows a behaviour consistent to the one we 
have discussed for $\zeta_2$. For both sets of exponents, it is clear that the measurements 
for different resolutions are consistent with each other within error-bars. 

Let us stress that the behaviour of equal-time exponents as a function of $b$
has been discussed before~\cite{mogensen,gilbert2002,ditlevsenbook}. However,
our systematic study does throw-up a few surprises which we cannot refrain from
commenting upon before we turn our attention to the dynamics of such systems.
Firstly, our detailed simulations show that the secondary scaling range yields 
and exponent which, within error-bars, is marginally larger than $\alpha$ as 
$b \to b_c$. Secondly, and more curiously, the two-dimensional turbulent behaviour -- a single 
scaling range for $n < n_f$ -- with $\zeta_p  = p/3$ is recovered only asymptotically 
as $b \to -2$.

\begin{figure*}
\begin{center}
\hspace*{-1.0cm}
\includegraphics[scale=0.11]{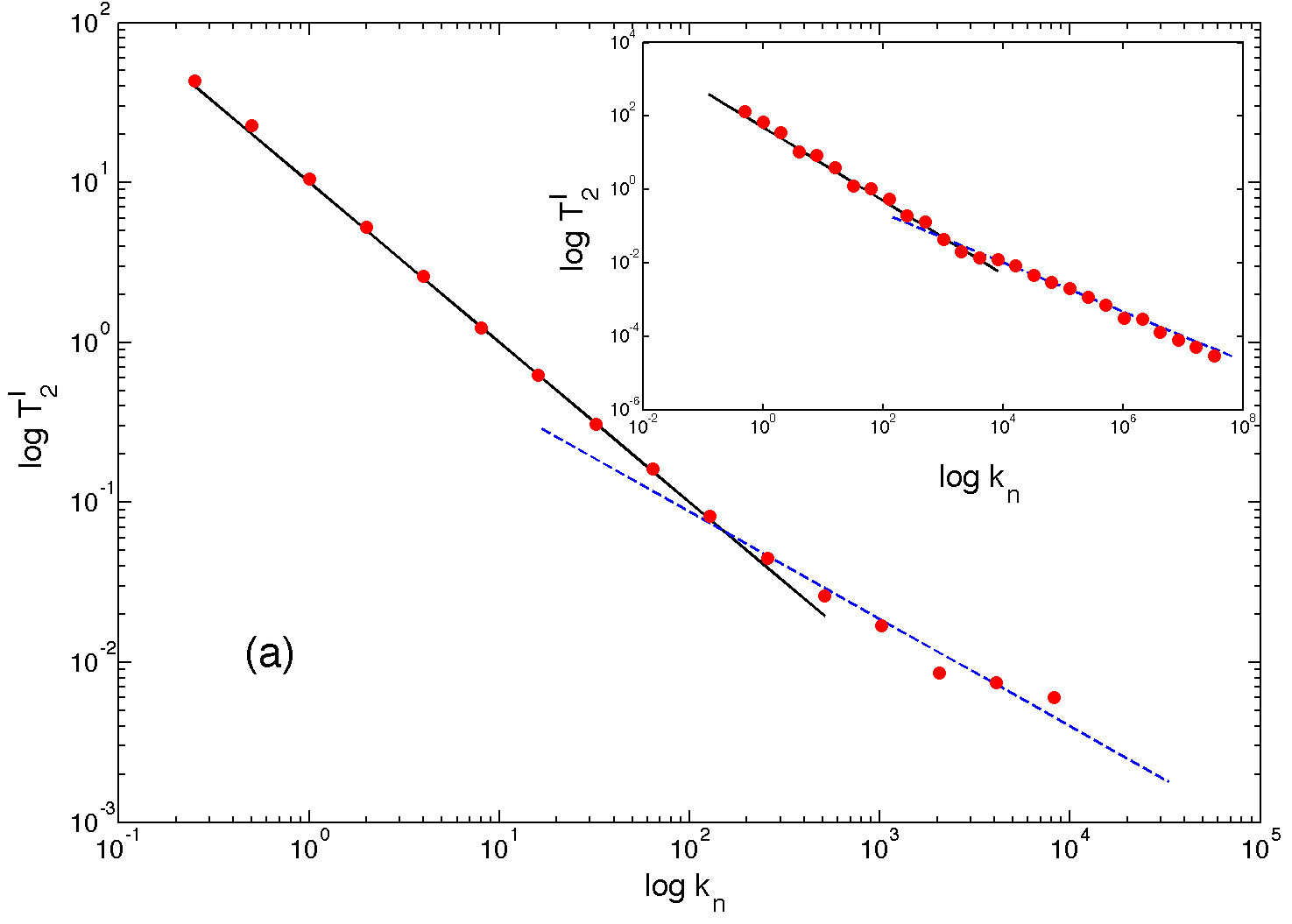}
\includegraphics[scale=0.11]{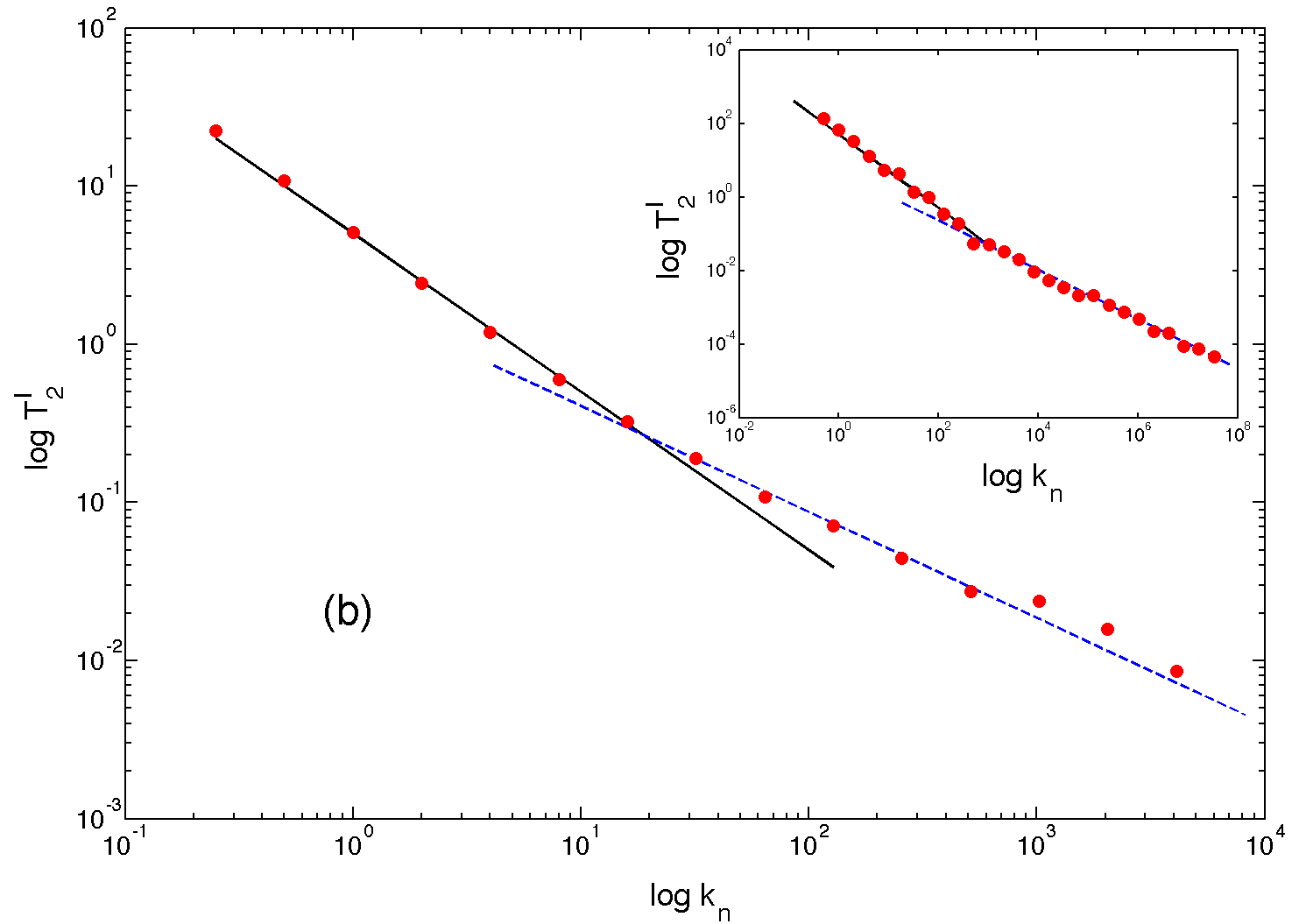}
\includegraphics[scale=0.11]{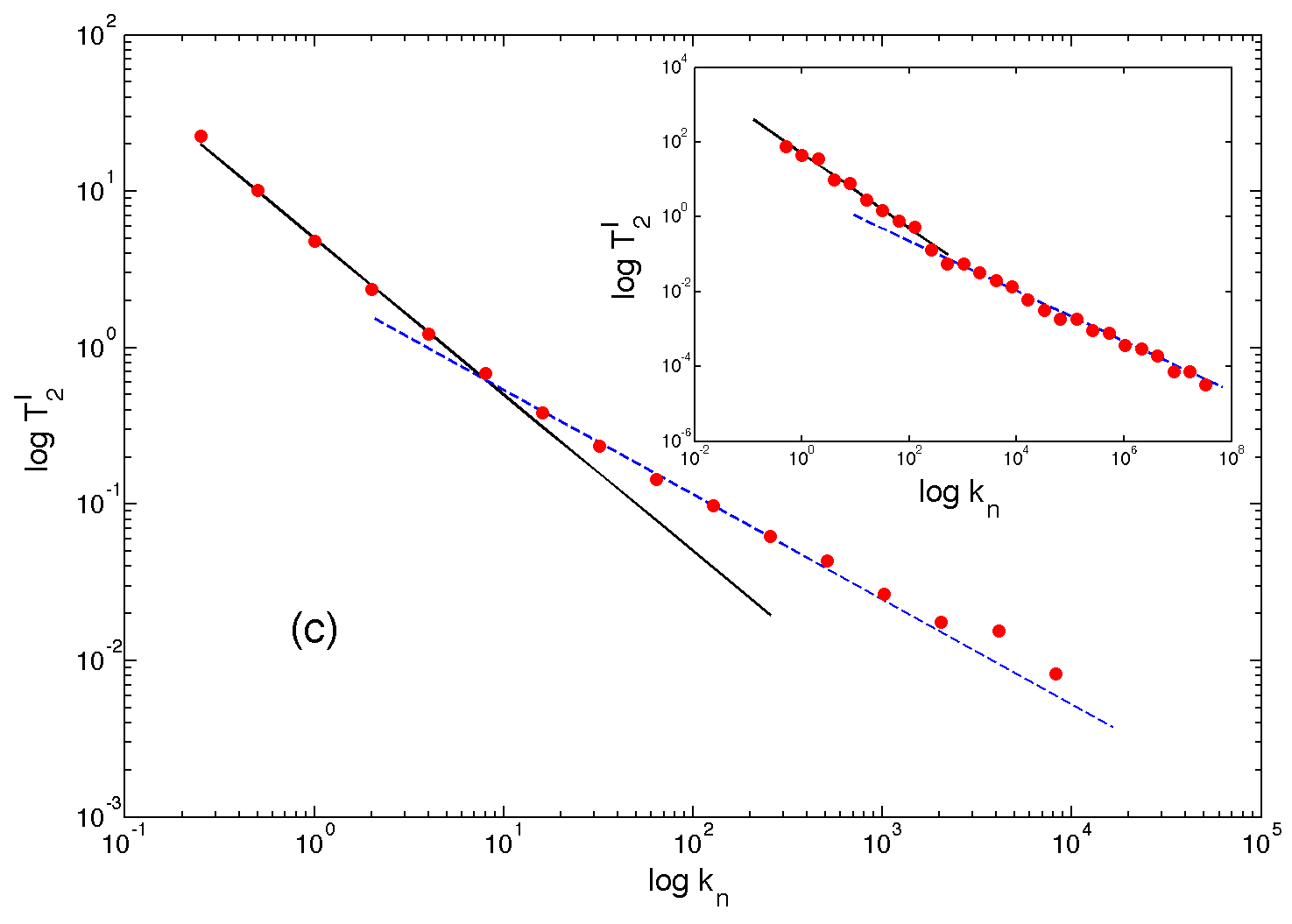}
\end{center}
\caption{(color online) Representative log-log plots, from runs {\bf R2} and {\bf R1} (in the insets), 
of the second-order integral time scale $T_2$ versus the wavenumber $k_n$ for  (a) $b = -1.4$, (b) $b = -1.55 $, and (c) $b = -1.7$. In panels (a) and 
(b), the black, dashed line 
corresponds to our theoretical prediction, from the energy equipartition arguments, of $k_n^{-1}$ ($z_2 = z_4 = 1$) and the blue, dashed line 
shows a scaling $k_n^{-\alpha/2 - 1}$ predicted from an enstrophy equipartition solution. In panel (c), where $b < b_c$, the  blue, dashed line 
indicates $k_n^{-2/3}$, as we would expect from a Kolmogorov-type, turbulent solution.}
\label{T2}
\end{figure*}

We finally turn our attention to the dynamics of such systems, especially near the 
critical point $b_c$. This is most conveniently done by examining the time scales associated 
with the time-dependent order-$p$ structure function, defined for the shell model as 
\begin{equation}
F_p(k_n,t) \equiv \la [u_n(t_0)u^{\ast}_n(t_0 + t)]^{p/2} \ra.
\label{goy2}
\end{equation}
This allows us to extract the integral-time scale~\cite{mitra04,mitra05,raynjp,rayepjb} via the time integral
\begin{equation}
T_p(k_n) = \frac{1}{F_p\vert_{t=0}} \int_0^{t_\epsilon} F_p ds
\label{int-time}
\end{equation}
and, thence, via the dynamic scaling {\it Ansatz} $T_p = k_n^{-z_p}$ the dynamic exponent $z_p$. In this 
integral the upper limit $t_\epsilon$ is taken as the time when the time-dependent structure function for a particular 
shell (Fig.~\ref{T2-asym}a (inset)) falls below a threshold $\epsilon$; for our calculations we choose $\epsilon = 0.6$ but 
have checked that our results are unchanged for $0.4 \le \epsilon \le 0.7$.

In Fig.~\ref{T2-asym} we show the representative plots of the second and fourth-order 
integral time scales, versus the wavenumber, extracted from time-dependent structure functions (inset of 
Fig.~\ref{T2-asym}a) for extremal values of the parameter (a) $b=-1.2$ and (b) $b=-1.9$. We see clearly that
between these two values the scaling behaviour switches from the equipartition $k_n^{-1}$ to the 
Kolmogorov scaling $k_n^{-2/3}$. In order to understand this transition clearly, it is important to 
examine the integral time-scales for intermediate values of $b$. 

\begin{figure}
\begin{center}
\includegraphics[scale=0.16]{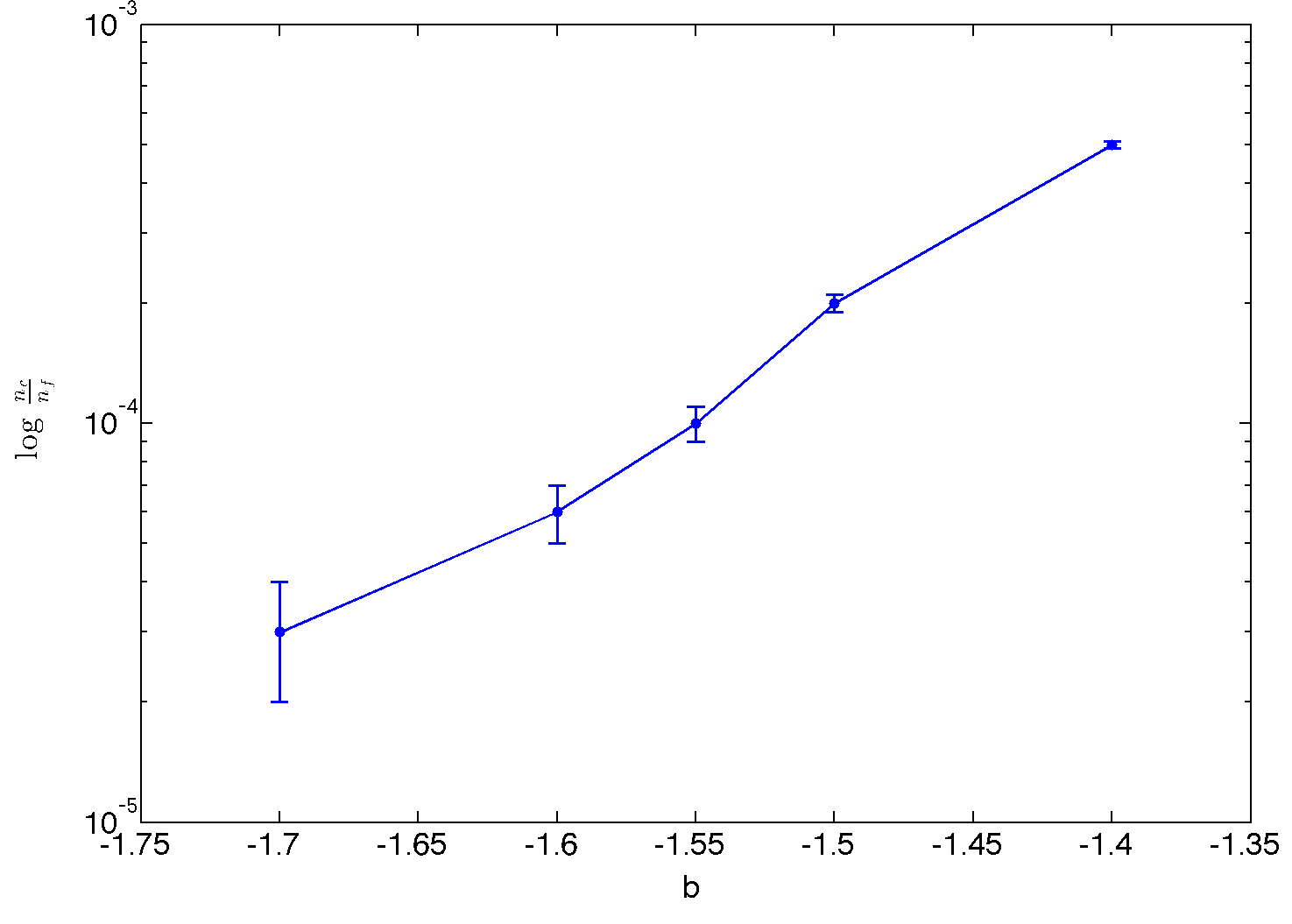}
\end{center}
\caption{(color online) A plot of the ratio of the estimated cross over shell $n_c$ with the forcing shell $n_f$ versus the parameter $b$ in the 
intermediate co-existence region. Within errorbars 
the curves for runs {\bf R1} and {\bf R2} agree with each other. We present here results from {\bf R1}.}
\label{nc}
\end{figure}

In Fig.~\ref{T2} we show representative, log-log plots of $T_2$ vs $k_n$ for $n
< n_f$ for such intermediate values of $b$. We see the clear co-existence of both scaling ranges 
as shown by the thick black line ($k_n^{-1}$) and the broken blue line ($k_n^{-2/3}$) threading the data. 
With decreasing $b$, the second scaling regime starts to dominate and at the cost of the first. The fourth-order 
structure functions show an identical behaviour and, hence, are not presented for brevity. It is possible 
to estimate the cross-over scales from such plots and in Fig.~\ref{nc} we show a plot of the ratio of the cross-over 
wavenumber to the forcing wavenumber as a function of $b$ (in the intermediate range). Given the discrete and exponential 
spacing of the shell space, we should keep in mind that such estimates are not likely to be very precise.

Our results for the dynamic exponents is interesting for several reasons. It
useful to remind ourselves that, by definition, dynamic exponents in shell
models are not affected by sweeping which would trivially yield $z_2 = 1$.
Within the multifractal model of three-dimensional turbulence, 
the dynamic exponents are related to the equal-time ones through a linear bridge relation.
Here, however, $z_2 = 1$ is the result of a unique
time scale $\tau = 1/k_n$ because of energy equipartition. Similarly, the second scaling regime is due to the
equipartition of the generalised enstrophy which yields, again, a unique time
scale $\tau \sim (u_nk_n)^{-1} = k_n^{-\alpha/2 - 1}$. For values of $b$
smaller than the critical value, both these scaling regimes disappear and we
end up with a Kolmogorov-like dynamic exponent $z = 2/3$ consistent with $\zeta_2 = 2/3$~\cite{raynjp}.
It is useful to draw
attention to the fact that the Kolmogorov-like dynamic exponent is, in its
numerical value, close to that obtained via an equipartition argument for
values of $b$ close to -2. Hence extreme care needs to be taken to disentangle
the two effects. It is for these reasons that we performed two sets of
simulations with very different extents of the scaling range and found the
exponents extracted from both these runs to be in agreement with each other. 

In this paper, in the light of recent work on decimated Navier-Stokes
turbulence, we revisited the two-dimensional SABRA model to understand, within
the simplifications of a shell model, the interplay between cascade and
equilibrium solutions. We ought to keep in mind that our work builds on and
adds to  the previous studies of
Refs.~\cite{mogensen,gilbert2002,ditlevsenbook}. In particular, we have
explored systematically the phase diagram proposed by Gilbert {\it et
al.}~\cite{gilbert2002} and found some surprises. Furthermore we have measured,
to our knowledge for the first time, the dynamics of such systems and their
associated time scales which, unsurprisingly, are very different from the
dynamic multiscaling that we usually associate with  fully developed
turbulence. Most curiously, our studies show, contrary to previous
estimates~\cite{ditlevsenbook}, that the SABRA model mimics two-dimensional
turbulence only asymptotically as $b\to -2$. These studies thus reemphasize the
hope~\cite{frisch12} that there might be more in common between statistical
mechanics and true two-dimensional turbulence than was thought before. It is
left for the future to evaluate the relaxation to equilibrium states and the
dynamics close to the critical dimension of Frisch {\it et al}.~\cite{frisch12}
in direct numerical simulations of the decimated, two-dimensional Navier-Stokes
equation.

The authors are grateful to Anna Pomyalov for many insightful discussions and
suggestions. 
SSR acknowledges the support of the DAE and the DST (India) project ECR/2015/000361. 
Our simulations were performed on the cluster {\it Mowgli}
and the work station {\it Goopy} at the ICTS-TIFR.


\begin{thebibliography}{10}
\bibitem{k41} A.N. Kolmogorov, Dokl. Akad. Nauk SSSR 
{\bf 30}, 301 (1941); A.N. Kolmogorov, Dokl. Akad. Nauk SSSR 
{\bf 31}, 538 (1941).
\bibitem{frischbook} U. Frisch, {\it Turbulence: The Legacy of A.N.
Kolmogorov} (Cambridge University, Cambridge, UK, 1996).
\bibitem{falcormp} G. Falkovich, K. Gawedzki and M. Vergassola, 
Rev. Mod. Phys. {\bf 73}, 913 (2001).
\bibitem{pandit-review} R. Pandit, P. Perlekar, and S.S. Ray, Pramana {\bf 73}, 157 (2009).
\bibitem{hopf} E. Hopf, {\it Comm. Pure Appl. Math.} {\bf 3}, 201 (1950).
\bibitem{lee} T. D. Lee,  {\it Q. J. Appl. Math.} {\bf 10} (1952).
H.K.~Moffatt and K.S.R.~Sreenivasan, eds., Cambridge University Press,
Cambridge, (2010).
\bibitem{brachet05} C. Cichowlas, P. Bona\"iti, F. Debbash, and M. Brachet,  {Phys. Rev. Lett.} {\bf 95},
264502 (2005).
\bibitem{ray11}  S. S. Ray, U. Frisch, S. Nazarenko, and T. Matsumoto, {Phys. Rev. E} {\bf 84}, 16301 (2011).
\bibitem{bannerjee14} D. Banerjee and S. S. Ray, Phys. Rev. E {\bf 90}, 041001(R) (2014).
\bibitem{ray-review} S. S. Ray, in {\it Persp. in Nonlinear Dynamics}, Pramana - J. of Phys. {\bf 84}, 395, (2015).
\bibitem{divya17} D. Venkataraman and S. S. Ray, Proc. Royal Soc. A {\bf 473}, 20160585 (2017).
\bibitem{frisch08} U. Frisch, S. Kurien, R. Pandit, W. Pauls, S. S. Ray, A. Wirth, and J-Z Zhu,  
{Phys. Rev. Lett.} {\bf 101}, 144501 (2008).
\bibitem{frisch13} U. Frisch, S. S. Ray, G. Sahoo, D. Banerjee, and R. Pandit,  
{Phys. Rev. Lett.}, {\bf 110}, 64501 (2013).
\bibitem{bottleneck} W. Dobler, N.E.L. Haugen, T.A. Yousef and A.
Brandenburg, {Phys. Rev. E}, {\bf 68}, 026304 (2003);
Z.-S. She, G. Doolen, R.H. Kraichnan, and S.A. Orszag, {Phys. Rev. Lett.}, {\bf 70}, 3251 (1993); P.K. Yeung and Y. Zhou, {Phys.
Rev. E}, {\bf 56}, 1746 (1997); T. Gotoh, D. Fukayama, and T. Nakano, {Phys.  Fluids}, {\bf 14}, 1065 (2002); M.K. Verma and 
D.A. Donzis, {J.  Phys. A: Math.  Theor.}, {\bf 40}, 4401 (2007); P.D. Mininni, A. Alexakis, and
A. Pouquet, {Phys. Rev. E}, {\bf 77}, 036306 (2008).
Y. Kaneda, {\it et al.},  {Phys. Fluids}, {\bf 15},
L21 (2003); T. Isihara, T. Gotoh, and Y. Kaneda {Annu. Rev.  Fluid Mech.},
{\bf 41}, 165 (2009); S. Kurien, M.A. Taylor, and T. Matsumoto, {Phys.
Rev. E}, {\bf 69}, 066313 (2004); D.A. Donzis and K.R. Sreenivasan {J. Fluid Mech.}, {\bf
657}, 171 (2010); H.K. Pak, W.I. Goldburg, A. Sirivat, {Fluid Dynamics
Research}, {\bf 8}, 19 (1991); Z.-S. She and E. Jackson, {Phys. Fluids A},
{\bf 5}, 1526 (1993); S.G. Saddoughi and S.V. Veeravalli, {J. Fluid Mech.},
{\bf 268}, 333 (1994); G. Falkovich, {Phys. Fluids}, {\bf 6}, 1411 (1994);
L. Sirovich, L. Smith, and V. Yakhot, {Phys. Rev. Lett.},
{\bf 72}, 344, (1994).
\bibitem{lvov02} V. L'vov, A. Pomyalov and I. Procaccia, {Phys. Rev. Lett.}  {\bf 89}, 064501 (2002).
\bibitem{frisch12} U. Frisch, A. Pomyalov, I. Procaccia, and S. S. Ray, {Phys. Rev. Lett.} {\bf 108}, 074501 (2012). 
\bibitem{luca15} A. S. Lanotte, R. Benzi, S. K. Malapaka, F.  Toschi, and L. Biferale, Phys. Rev. Lett. {\bf 115}, 264502 (2015).
\bibitem{luca16} A. S. Lanotte, S. K. Malapaka, L. Biferale, Eur.  Phys. J. E {\bf 39}, 49 (2016).
\bibitem{michele17} M. Buzzicotti, A. Bhatnagar, L. Biferale, A. S. Lanotte, S. S. Ray, New J. of Phys. {\bf 18}, 113047 (2016).
\bibitem{michele16} M. Buzzicotti, L. Biferale, U. Frisch, and S. S. Ray, Phys. Rev. E {\bf 93}, 033109 (2016).
\bibitem{mitra04} D. Mitra and R. Pandit, Phys. Rev. Lett. {\bf 93}, 024501 (2004).
\bibitem{mitra05} D. Mitra and R. Pandit, Phys. Rev. Lett. {\bf 95}, 144501 (2005).
\bibitem{rayepjb} R. Pandit, S. S. Ray and D. Mitra, Eur. Phys. J. B {\bf 64}, 463 (2008).
\bibitem{raynjp}  S. S. Ray, D. Mitra and R. Pandit, New J. Phys. {\bf 10}, 033003 (2008).
\bibitem{luca11} L. Biferale, E. Calzavarini, and F. Toschi, Phys. Fluids {\bf 23}, 085107 (2011).
\bibitem{rayprl11} S. S. Ray, D. Mitra, P. Perlekar, and R. Pandit, Phys. Rev. Lett. {\bf 107}, 184503 (2011).
\bibitem{parisifrisch} G. Parisi and U. Frisch in {\it Turbulence and
Predictability of Geophysical Fluid Dynamics}, eds. M. Ghil, R. Benzi, and
G. Parisi (North-Holland, Amsterdam, 1985) p 84.
\bibitem{mogensen} P. D. Ditlevsen and I. A. Mogensen, Phys. Rev.  E {\bf 53}, 4785 (1996).
\bibitem{gilbert2002} T. Gilbert, V. S. L'vov, A. Pomyalov, and I. Procaccia, Phys. Rev. Lett. {\bf 89} 074501 (2002). 
\bibitem{luca-review} L. Biferale, Annu. Rev. Fluid Mech. {\bf 35}, 441 (2003)
\bibitem{sabra} V.S. L'vov, E.Podivilov, A. Pomyalov, I. Procaccia, and D. Vandembroucq, Phy. Rev. {\bf 58}, 1811 (1998)
\bibitem{goy} E. B. Gledzer, Dokl. Akad. Nauk SSSR {\bf 209}, 1046 (1973)
[Sov. Phys. Dokl. {\bf 18}, 216 (1973)]; M. Yamada and K. Ohkitani, Phys. Rev. Lett. {\bf 60}, 983
(1988).
\bibitem{ditlevsenbook} P. D. Ditlevsen, {\it Turbulence and Shell Models}
(Cambridge University, Cambridge, UK, 2010).
\bibitem{bohrbook} T. Bohr, M. H. Jensen, G. Paladin and A. Vulpiani, {\it Dynamical systems approach to
turbulence} (Cambridge University Press, Cambridge, UK, 1998).
\bibitem{foot1} We note that this is
in the spirit of standard DNSs of two-dimensional Navier-Stokes
equation~\cite{rayprl11,perlekar11} where a friction term is added to mimic
air-drag-induced friction which prevents accumulation of energy at the largest
scales.  
\bibitem{perlekar11} P. Perlekar, S. S. Ray, D. Mitra, and R. Pandit, Phys. Rev. Lett. {\bf 106}, 054501 (2011).
\bibitem{pisarenko}D. Pisarenko, L. Biferale, D. Courvoisier, 
U. Frisch, and M. Vergassola, Phys. Fluids A {\bf 5}, 2533 (1993).
\bibitem{dhar}S. Dhar, A. Sain, and R. Pandit, Phys. Rev. Lett. 
{\bf 78}, 2964 (1997).

\end{thebibliography}
\end{document}